\documentclass[fleqn,10pt]{wlscirep}
\usepackage{color}
\usepackage{siunitx}
\usepackage{float}
\usepackage{xspace}
\newcommand{\singlecolumn}{89mm}
\newcommand{\doublecolumn}{183mm}

\newcommand{\binomialeq}[2]{C^{#1}_{#2}} % 
\newcommand{\ket}[1]{\left| #1 \right>} % for Dirac bras
\newcommand{\STp}{S-T$_+$\xspace} % S-T+

\newcommand{\Tcoherence}{$T_\text{coh}$\xspace} % T2*
\newcommand{\Tcoherenceeq}{T_\text{coh}} % T2*

\newcommand{\Ndots}{$N_\text{dots}$\xspace} % T2*
\newcommand{\Ndotseq}{N_\text{dots}} % T2*

\newcommand{\VTLeq}{V_\text{TL}} % V_TL
\newcommand{\VTeq}{V_\text{T}} % V_T
\newcommand{\VTTLeq}{V_\text{T-TL}} % V_T-TL
\newcommand{\VTTReq}{V_\text{T-TR}} % V_T-TR
\newcommand{\VBeq}{V_\text{B}} % V_B
\newcommand{\VBBLeq}{V_\text{B-BL}} % V_B-BL
\newcommand{\VBBReq}{V_\text{B-BR}} % V_B-BR
\newcommand{\VReq}{V_\text{R}} % V_R
\newcommand{\VRBReq}{V_\text{R-BR}} % V_R-BR
\newcommand{\VRTReq}{V_\text{R-TR}} % V_R-TR
\newcommand{\VLeq}{V_\text{L}} % V_L
\newcommand{\VLTLeq}{V_\text{L-TL}} % V_L-TL
\newcommand{\VLBLeq}{V_\text{L-BL}} % V_L-BL
\newcommand{\VXeq}{V_\text{X}} % V_X
\newcommand{\VYeq}{V_\text{Y}} % V_Y

\newcommand{\dVLTLeq}{\delta \VLTLeq} % V_L-TL
\newcommand{\dVTeq}{\delta \VTeq} % V_T
\newcommand{\dVBeq}{\delta \VBeq} % V_B
\newcommand{\dVReq}{\delta \VReq} % V_R
\newcommand{\dVLeq}{\delta \VLeq} % V_L
\newcommand{\dVXeq}{\delta \VXeq} % V_X
\newcommand{\dVYeq}{\delta \VYeq} % V_Y

\newcommand{\VTL}{$\VTLeq$\xspace} % V_TL
\newcommand{\VT}{$\VTeq$\xspace} % V_T
\newcommand{\VTTL}{$\VTTLeq$\xspace} % V_T-TL
\newcommand{\VTTR}{$\VTTReq$\xspace} % V_T-TR
\newcommand{\VB}{$\VBeq$\xspace} % V_B
\newcommand{\VBBL}{$\VBBLeq$\xspace} % V_B-BL
\newcommand{\VBBR}{$\VBBReq$\xspace} % V_B-BR
\newcommand{\VR}{$\VReq$\xspace} % V_R
\newcommand{\VRBR}{$\VRBReq$\xspace} % V_R-BR
\newcommand{\VRTR}{$\VRTReq$\xspace} % V_R-TR
\newcommand{\VL}{$\VLeq$\xspace} % V_L
\newcommand{\VLTL}{$\VLTLeq$\xspace} % V_L-TL
\newcommand{\VLBL}{$\VLBLeq$\xspace} % V_L-BL
\newcommand{\dVX}{$\delta \VXeq$\xspace} % V_X
\newcommand{\dVY}{$\delta \VYeq$\xspace} % V_Y

\newcommand{\depsilonLTLeq}{\delta V_\text{L-TL}^\epsilon} % V_R
\newcommand{\dJLTLeq}{\delta V_\text{L-TL}^J} % V_L
\newcommand{\depsilonCReq}{\delta V_\text{C-R}^\epsilon} % V_X
\newcommand{\dJCReq}{\delta V_\text{C-R}^J} % V_Y
\newcommand{\depsilonLTL}{$\depsilonLTLeq$\xspace} % V_R
\newcommand{\dJLTL}{$\dJLTLeq$\xspace} % V_L
\newcommand{\Lone}{$\text{L}_1$\xspace} % L_1
\newcommand{\LS}{$\text{L}_\text{S}$\xspace} % L_S
\newcommand{\Isolated}{$\text{I}$\xspace} % I
\newcommand{\Readout}{$\text{R}$\xspace} % R

\title{Coherent control of individual electron spins in a two dimensional array of quantum dots}

\author[1]{Pierre-Andr\'e Mortemousque}
\author[1]{Emmanuel Chanrion}
\author[1]{Baptiste Jadot}
\author[1]{Hanno Flentje}
\author[2]{Arne Ludwig}
\author[2]{Andreas D. Wieck}
\author[1]{Matias Urdampilleta}    
\author[1]{Christopher B{\"a}uerle}    
\author[1]{Tristan Meunier}

\affil[1]{Univ. Grenoble Alpes, CNRS, Grenoble INP, Institut N\'eel, 38000 Grenoble, France}%
\affil[2]{Lehrstuhl f{\"u}r Angewandte Festk{\"o}rperphysik, Ruhr-Universit{\"a}t Bochum, Universit{\"a}tsstra{\ss}e 150, D-44780 Bochum, Germany}%

\begin{abstract}
The ability to manipulate coherently individual quantum objects organized in arrays is a prerequisite to any scalable quantum information platform. For electron spin qubits, it requires the fine tuning of large arrays of tunnel-coupled quantum dots.
The cumulated efforts in linear dot arrays have permitted the recent realization of quantum simulators and multi-electron spin coherent manipulation. 
However, the two-dimensional scaling of such implementations remains undemonstrated while being compulsory to resolve complex quantum matter problems or process quantum information. 
Here, we demonstrate the two-dimensional coherent control of individual electron spins in a $3\times 3$ array of tunnel coupled quantum dots. 
More specifically, we focus on several key quantum functionalities of such control: charge deterministic displacement, local spin readout, local coherent exchange manipulation between two electron spins trapped in adjacent dots, and coherent multi-directional spin shuttling over distances of several microns. 
This work lays the foundations for exploiting a two-dimensional array of electron spins for quantum simulation and information processing.
\end{abstract}
\begin{document}

%%%%%%%%%%%%%%%%%%%%%%%%%%%%%%%%%%%%%%%%%%%%
% BEGIN
%%%%%%%%%%%%%%%%%%%%%%%%%%%%%%%%%%%%%%%%%%%%
\flushbottom
\maketitle
% ^ <matias.urdampilleta@gmail.com> 2018-06-05T13:13:06.071Z.
\thispagestyle{empty}

\section*{Introduction}
%\bfseries
A natural way to address the scalability of quantum devices is to design and realize two dimensional (2D) arrays of individual quantum objects with nearest neighbor interaction.\cite{Shor1995, Bacon2006, Fowler2011, Fowler2012} 
Increasing the size of the array is the subject of an intense research activity in many different quantum systems with the common objective to simulate the complex many-body problem \cite{dagotto1996surprises} and explore quantum coherence at large scale. 
It is also an important intermediate step towards the hypothetical realization of quantum information processors.\cite{Bacon2006, Fowler2011, Fowler2012} 
In large-scale semiconductor quantum processors, the quantum bit is defined by the spin of electrons trapped in an array of quantum dots (QDs).\cite{Veldhorst2017, Vandersypen2017} 
Over the years, an increasing number of QDs have been successfully controlled with demonstrations of individual spin quantum manipulation\cite{Ito2016, Fujita2017} and simulation of condensed matter problems such as the Fermi-Hubbard model.\cite{Hensgens2017} 
However, all these experiments have been performed in linear arrays of tunnel-coupled quantum dots. 
Important technological challenges remain to be addressed before reaching the required level of control in 2D quantum manipulation.\cite{Thalineau2012, Mukhopadhyay2018} 
In particular, identifying the gate layout of a 2D dot array compatible with a planar geometry, controlling the electron filling of arrays and engineering the coherent control and readout of spins are among the basic procedures still lacking for the control of 2D QD arrays. 

In this article, we report on the coherent control of individual electrons in a $3\times 3$ array of tunnel coupled quantum dots. 
Thanks to the recent developments on isolated dot systems,\cite{Bertrand2015, Flentje2017, Flentje2017-2} we fix the number of electrons loaded in the QD array. 
After tuning the system in the moderate tunnel coupling regime, all possible charge configurations of the QD array are identified. 
We then implement several key procedures for the control of electron spins in quantum dot arrays. 
Firstly, the two-electron spin readout at any location in the quantum dot array is demonstrated. 
Secondly, the local enhancement of the tunnel coupling is implemented to reach the regime where the coherent exchange of a quantum of spin between two electrons sitting in adjacent quantum dots is observable. 
This key control allows the observation of spin exchange oscillations at the detuning sweet spot. 
Finally, coherent displacements within the array are investigated by separating two electrons which are initially prepared in the singlet state. 
We demonstrate one- and two-spin coherent displacement in different directions as well as a hyperbolic increase of the spin coherence time with the electron speed. 
All these results demonstrate an unprecedented control of electron spins in 2D arrays, with demonstration of quantum functionalities on individual electron spins crucial in the prospect of quantum simulation and quantum computation.   
%\mdseries

%%%%%%%%%%%%%%%%%%%%%%%%%%%%%%%%%
%%%%%%%%%%%%%%%%%%%%%%%%%%%%%%%%%
% Results
%%%%%%%%%%%%%%%%%%%%%%%%%%%%%%%%%
%%%%%%%%%%%%%%%%%%%%%%%%%%%%%%%%%
\section*{Results}
\begin{figure}[H]
\centering
\includegraphics[width=\doublecolumn]{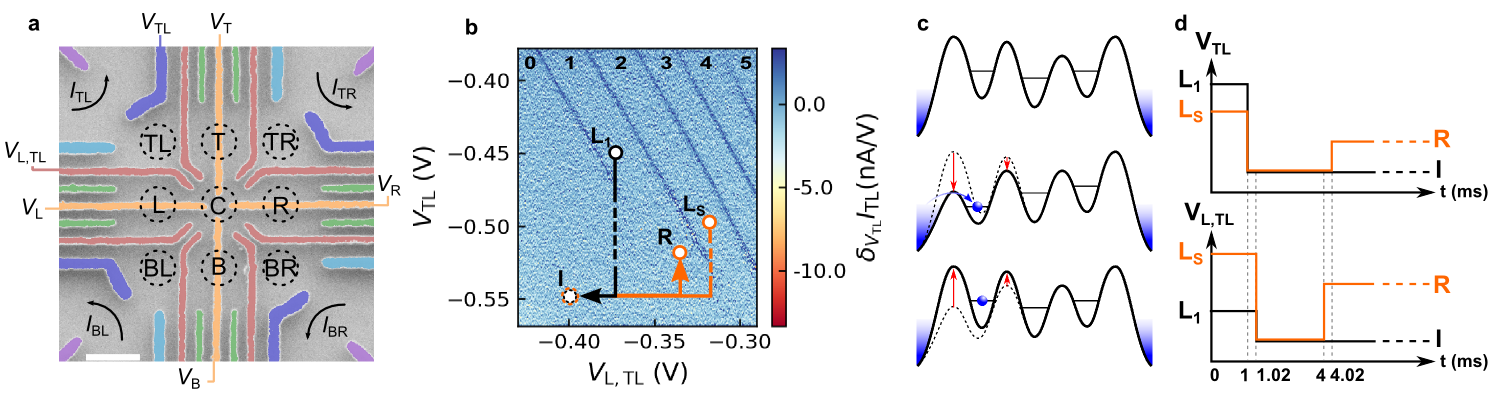}
\caption{
\textbf{Two-dimensional array of quantum dots in the isolated regime.}  
\textbf{a,}~Electron micrograph of a sample similar to the one used in this work. 
The nine \textit{dashed circles} indicate the $3\times3$ array of quantum dots (QDs) arising from the  gate-induced potential landscape (see text). 
The \textit{purple} gates are used to define four quantum point contacts used as local electrometers, whose conductances set the measured currents $I_\text{TL, BL, BR, TR}$. 
\textit{Scale bar} (white) is $200$~\si{\nano\meter}. 
\textbf{b,}~Typical charge stability diagram of the TL QD. 
The derivative $\partial_{V_\text{TL}} I_\text{TL}$ of the recorded current is plotted as a function of the sweeping voltage \VTL and \VLTL. 
The electron occupation number is indicated at the top. 
The gate voltage configurations \Lone, \LS, \Isolated, and \Readout are respectively employed to load one electron, two electrons in the singlet spin state, to isolate the QD system from the electron reservoirs and to perform the spin readout (see Methods). 
\textbf{c,}~Schematics of a potential cut along the top line of QDs in the isolated regime I empty of electron (top), at the loading position \Lone (middle) and back at the isolated position \Isolated with a single electron loaded (bottom).
\textbf{d,}~Schematics of the pulse sequence applied on \VTL (top) and \VLTL (bottom) used to load electrons and bring them in the isolated configuration. The black and orange curves correspond respectively to the one- and two-electron pulse sequences.
}
\label{fig:sample}
\end{figure}

The 2D up-scaling of tunnel-coupled QDs requires a complex geometry of gates. For a 9 dot array, the simultaneous tuning of 28 different gate voltages was necessary (see Fig.~\ref{fig:sample}a). 
The outer edges of the dot array are defined by the \textit{green}, \textit{light blue} and \textit{blue} gates. 
The inner square lattice is tuned by adjusting the red gate voltages. 
The couplings of the corner QDs TL, BL, BR and TR with the electron reservoirs are controlled by the \textit{light blue} and \textit{blue} gates, and finally, the different QD potentials are tuned using the \textit{blue} and \textit{orange} gates independently, or linear combinations of  $V_\text{L, B, R, T}$. 
As an example, we define two \textit{virtual} gates,\cite{Hensgens2017} \dVX and \dVY, to act similarly to the in-plane $x$- and $y$- electric dipoles of the sample:
$\dVXeq(\alpha_\text{L}) = (\dVReq,-\alpha_\text{L}\dVLeq)$ and 
$\dVYeq(\alpha_\text{B}) = (\dVTeq,-\alpha_\text{B}\dVBeq)$, 
where $\alpha_\text{L}$ and $\alpha_\text{B}$ are specifically adjusted for different gate configurations to excite the two dipoles.

To investigate the control of the electron filling in the dot array, the first step consists of loading the dot structure with a finite number of electrons\cite{Bertrand2015, Flentje2017, Flentje2017-2} via the TL dot. 
Then, the array of QDs is isolated from the electron reservoir. 
The electron loading and isolation sequence is overlaid on the TL QD charge stability diagram shown in Fig.~\ref{fig:sample}b, and explained in the following: 
the interruption of the degeneracy lines of the 0-1 and 1-2 electrons at $\VTLeq = -0.47$~\si{\volt} and $\VTLeq = -0.53$~\si{\volt}, respectively, indicates that the tunnel barrier to the electron reservoir becomes thicker and thicker as \VTL is more negative. 
It is therefore possible, starting from a QD structure empty of electrons, to apply a voltage pulse on \VLTL and \VTL to load either one (\Lone, \textit{black} sequence) or two electrons (\LS, \textit{orange} sequence). 
A rapid pulse back to the point I brings the QD system in a regime isolated from the electron reservoir. 
Figure~\ref{fig:sample}c, d shows the time frame of the potential cut along the top row of QDs and corresponding \VLTL and \VTL voltage pulse sequence.

%%%%%%%%%%%%%%%%%%%%%%%%%%%%%%%%%
% Charge
\subsection*{Charge control of a single electron}

\begin{figure}[H]
\centering
\includegraphics[width=\singlecolumn]{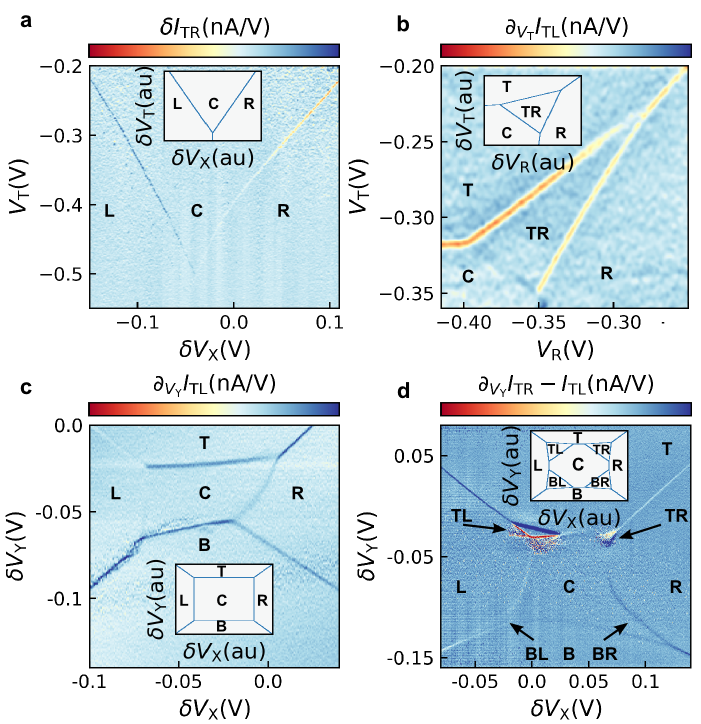}
\caption{
\textbf{Single electron charge configurations in the $3\times 3$ array of quantum dots.} 
\textbf{a,}~Charge stability diagram of one electron in the linear triple (L, C, R) quantum dot (QD) system. 
The derivative $\partial_{V_\text{T}} I_\text{TR}$ of the recorded current is plotted as a function of $V_\text{T}$ and $\delta V_\text{X}(1)$ (see text). 
The labels indicate the position of the isolated electron in the QD array. 
\textbf{b,}~Charge stability diagram of one electron in the four top right (T, TR, R, C) QDs. 
The derivative $\partial_{V_\text{T}} I_\text{TL}$ of the recorded current is plotted as a function of $V_\text{T}$ and $V_\text{TR}$. 
\textbf{c,}~Charge stability diagram of one electron in five QDs (L, B, R, T, C) in a `cross' geometry. 
The derivative $\partial_{V_\text{Y}} I_\text{TL}$ of the recorded current is plotted as a function of $\delta V_\text{Y}(0.64)$ and $\delta V_\text{X}(1)$. 
\textbf{d,}~Charge stability diagram of one electron in the $3 \times 3$ array of QDs. 
The current derivative $\partial_{V_\text{Y}} \left(I_\text{TR} - I_\text{TL} \right)$ is plotted as a function of $\delta V_\text{Y}(0.62)$ and $\delta V_\text{X}(0.97)$. 
The static potential landscape is tuned in slightly different configurations for all four experiments. 
The figure insets show stability diagram simulations (see Methods). 
The qualitative agreement between the experiment and the simulation topologies allows the charge configuration assignment. The virtual gates used to characterize the QD system are particularly suitable to distinguish the five charge configurations of the quintuple dot in the cross geometry. The last four charge configurations of the nine-dot structure with one loaded electron are not straightforwardly accessible with the choice of virtual gates. Nevertheless, they are anticipated with the results on the four quantum dots previously described and are observed as triangular regions around each triple point of the five-dot stability diagram. 
}
\label{fig:stabdiag}
\end{figure}

In the following paragraph, we describe the systematic investigation of increasingly larger subsets of QDs (Fig.~\ref{fig:stabdiag}) explored by a single electron in the QD array.
We start with the charge stability of a linear triple QD system (L, C, R). 
The corresponding diagram is shown in Fig.~\ref{fig:stabdiag}a. 
In this experiment, \VT and \dVX are used to control the potential of the C dot and the energy detuning between L and R, respectively. 
Three different charge sectors are identified corresponding to the three possible charge configurations of the triple dot system. 
The topology of the charge stability diagram is in excellent agreement with previously measured triple QDs.\cite{Flentje2017} 
It is important to note that, in the case of an array not decoupled from the electron reservoir, other charge configurations with different total charge numbers would be observed.\cite{Thalineau2012} 
As a result, keeping the total number of electrons constant allows a drastic simplification of the charge stability diagrams, and thus of the charge manipulation. 
%The gradient of the detected current is plotted against $V_\text{T}$, used to control the potential of C $\epsilon_\text{C}$, and $\delta V_X$, a virtual gate controlling the energy detuning $\epsilon_\text{L} -  \epsilon_\text{R}$ and defined so that $(\delta V_L, \delta V_R) = (\alpha \delta V_X, \beta \delta V_X)$.  
Next, we probe the charge stability of four QDs in a square configuration, using \VT and \VR (Fig.~\ref{fig:stabdiag}b). 
Similarly with the previous case, we can clearly identify the four expected charge sectors. 
%In the top region, where $V_\text{T}$ is maximal, the $\epsilon_\text{T}$ is minimal and the electron is in T. 
%From the top right triple point to higher values of $V_\text{T}$ and $V_\text{R}$, one can notice the charge degeneracy line corresponding to the tunneling of the electron between the T and R QDs.
%In the bottom left region where $V_\text{T}$ and $V_\text{R}$ are minimal, the T and R, as well as TR, QDs are depleted. This is possible because the gate (g) to dot (d) capacitance elements $C(\text{g,d})$) obeys $C(V_\text{T},\text{T}) > C(V_\text{T},\text{TR}) > C(V_\text{T},\text{C})$, and the symmetric relation. 
%Finally, the region for which the electron lies in the TR QD is delimited by the three triple points $\text{P}_1$, $\text{P}_2$, and $\text{P}_3$. 
Similar topologies of charge stability regions are observed for the three other rotational permutations (see Extended Data Fig.~\ref{supfig:symmstabdiag}). % in the Extended Data Fig.~\ref{supfig:stabdiagothertopologies}. 
%We found that the control of a larger number of QDs is challenging using only two physical gate voltages. 
%Therefore, following the work of Hensgens \textit{et al.} \cite{Hensgens2017}, we define a set of virtual gates $V_\text{V} = (-a_0^\text{T} - a_1^\text{T} \,V_\text{T} , - a_0^\text{B} + a_1^\text{B} \,V_\text{B})$ and $V_\text{H} = (-a_0^\text{L} - a_1^\text{L} \,V_\text{L} , - a_0^\text{R} + a_1^\text{R} \,V_\text{R})$, with positive coefficients $a$, that are more convenient to fully explore the charge configurations. 
%At the zero-th order, this couple of virtual gates allows the tuning of position of the minimum of potential in the 2D array of QDs. 
We now explore the different charge configurations of a single electron in five QDs in a cross geometry, and in the $3\times3$ array of nine QDs, shown in Fig.~\ref{fig:stabdiag}c and Fig.~\ref{fig:stabdiag}d, respectively.
%is performed using $\delta V_X$ and $\delta V_Y$, for slightly different electrostatic tuning of the array. 
%Linear combinations of the detected current gradient is shown in Fig.~\ref{fig:stabdiag}(c) (five QDs), and in Fig.~\ref{fig:stabdiag}(d) (nine QDs). 
The qualitative comparison of the experimental data with theoretical simulations (respectively shown in the insets) allows the accurate assignment of the different charge configurations. 
This demonstrates the ability to deterministically control a single electron in a two-dimensional $3\times3$ array of QDs. 
In the following, we will demonstrate that when two electrons are loaded, the measurement of their spin at different positions in the array is also an efficient probe of the spatial charge configuration. 
%With this choice, the charge stability diagram of five QDs matches the physical disposition in a \textit{plus} configuration. 
%In Fig.~\ref{fig:stabdiag}(b), the top (bottom) region with maximal (minimal) values of $V_\text{V}$ corresponds to the electron in the T (B) QD. 
%Similarly, the right (left) region with maximal (minimal) values of $V_\text{H}$ corresponds to the electron in the R (L) QD. 
%In the central region delimited by four triple points, the electron is in the C QD. 
%By combining the use of such virtual gates [Fig.~\ref{fig:stabdiag}(a)] and the proper tuning of $V_\text{TL}$, $V_\text{BL}$, $V_\text{BR}$ and $V_\text{TR}$ [Fig.~\ref{fig:stabdiag}(b)], one can obtain the charge stability diagram of the $3 \times 3$ array of QDs for one electron [see Fig.~\ref{fig:stabdiag}(c)]. 

%%%%%%%%%%%%%%%%%%%%%%%%%%%%%%%%%
% Spin

\subsection*{Spin initialization, readout and displacement of two electron spins within the array}
% \subsection*{Spin initialization, readout and displacement of a two-electron spin state within the array}

\begin{figure}[H]
\centering
\includegraphics[width=\singlecolumn]{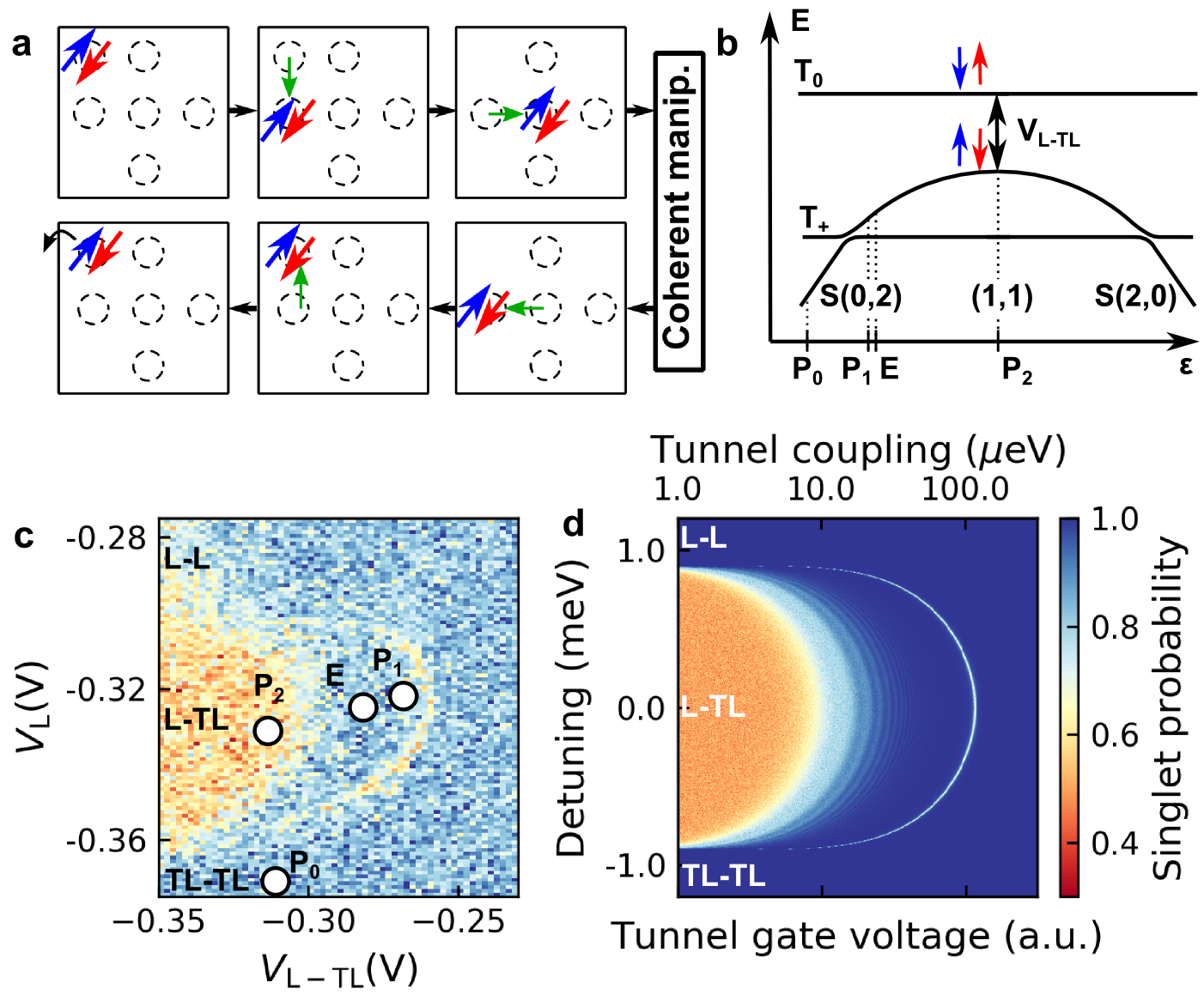}
\caption{
\textbf{Spin initialization, readout and manipulation in the quantum dot array.} 
\textbf{a,}~Top, electron spin manipulations used to initialize a singlet state in C: two-electron loading and spin relaxation in TL (left), coherent displacement in L (middle), and then in C (right).
Bottom, symmetric sequence executed to read out the final spin state: coherent displacement to L (right) and then TL (middle), and spin readout (left) using the TL electron reservoir (see Extended Data Fig.~\ref{supfig:spinreadout}). 
\textbf{b,}~Energy diagram of the relevant spin states for two electrons in the TL-L double dot as a function of the detuning between the two dots. When the two electrons are either in L or TL dot, the singlet state is the ground state respectively indicated as S(0,2) and S(2,0). The lowest energy state in the separated configuration is the triplet state $\text{T}_+$, where the spins are aligned and which crosses the singlet branch in two detuning points. At zero detuning and low tunnel coupling, the singlet state and the triplet $\text{T}_{0}$ with antiparallel spins mix together as a result of the different Zeeman energy in the two dots.  
\textbf{c,}~Experimental spin-mixing map of the double quantum dot L-TL performed under a $120$~\si{\milli\tesla} external magnetic field (see Methods). The singlet probability is plotted as a function of the voltages \VL and \VLTL. Two electrons in the singlet spin state initially located in TL are displaced over 50~\si{\nano\second} by a voltage pulse simultaneously applied on both \VL and \VLTL. 
\textbf{d,}~Simulated spin-mixing map realized using a homogeneous charging energy of $1$~\si{\milli\electronvolt}, an on-site spin exchange energy of $100$~\si{\micro\electronvolt} and a Zeeman energy of $3$~\si{\micro\electronvolt}. 
%
%\textbf{e, } Coherent exchange oscillations realized with a spin singlet initially in TL, adiabatically transferred to the L-TL charge configuration in the spin $\ket{\uparrow\downarrow}$ (voltage pulse from $\text{P}_0$ to $\text{P}_1$, then slowly ramped to $\text{P}_2$, see \textbf{b}), and rapidly pulse to E for a duration $\tau_E$.\cite{Petta2005} 
%
%\textbf{f, } Schematic $\VLTLeq(t)$ and $\VLeq(t)$ voltage sequences applied to perform the coherent exchange oscillations in \textbf{d}.\\
%
}
\label{fig:spininitreadout}
\end{figure}

To explore the spin dynamics of two electrons inside the 2D array of QDs, it is compulsory to initialize and read out the spin state of the electrons at any position in the array. 
Our implementation of such functionalities requires i) spin initialization, ii) spin preserving transfer of the electrons through the dot array, and iii) spin readout.\cite{Baart2016-3, Bertrand2016,  Flentje2017-2} Both initialization and readout are performed in the TL dot. 
The initialization procedure is presented in Fig.~\ref{fig:spininitreadout}a and consists of loading two electrons, and let the two-electron spin relax to the spin singlet ground state for $t_{init}=1\,\si{\milli\second} \gg \text{T}_1$ at the point \LS in Fig.~\ref{fig:sample}b.\cite{Petta2005} 
The spin readout is performed by measuring the singlet probability of the two-electron spin using a single-shot tunnel-rate selective spin readout (point \Readout in Fig.~\ref{fig:sample}b).\cite{Flentje2017-2, Hanson2005, Meunier2007} 
To read out the spin states at any dot location, the electrons are brought together to the TL dot with a spin-preserving transfer procedure. 
Considering the characteristics of the dot array, we demonstrate that the spin states can be transferred in two configurations, either between TL and L dots or within the L-B-R-T-C cross bar geometry.  

In order to achieve spin-preserving electron displacement, it is important to control the tunnel barrier between the TL and L dots using \VLTL (see Methods). Indeed, the tunnel barrier controls the amplitude of the exchange interaction as illustrated in Fig.~\ref{fig:spininitreadout}b. 
For large \VLTL, the exchange interaction is large enough to keep the singlet unaltered at any detuning position, except at the \STp avoided crossing where mixing can occur. 
For lower transfer value of \VLTL, and at low detuning, the exchange interaction can be dominated by the hyperfine coupling with the nuclear spins from the substrate. As a result, spin mixing can occur and the singlet is lost. 
Figure~\ref{fig:spininitreadout}c represents the spin-mixing map where the singlet probability is plotted as a function of \VLTL and detuning between the TL and L dots (the external magnetic field is set to 120~\si{\milli\tesla}). 
In this experiment, the electron spins are initialized in singlet in TL, and the tunnel coupling between TL and L is tuned using \VLTL ($x$-axis). 
Then, a 50-\si{\nano\second} long voltage pulse \VL ($y$-axis) changes the energy detuning between TL and L. 
Finally, the spin readout is performed at the point \Readout in Fig.~\ref{fig:sample}b. 
The spin map shows that the two-electron spin singlet is preserved when the exchange interaction is dominant, i.e. when the two electrons remain in the same quantum dot ($\VLeq > -0.3$~\si{\volt} or $\VLeq < -0.36$~\si{\volt}), or when they are separated in a highly-coupled double QD ($\VLTLeq > -0.3$). 
On the contrary, singlet-triplet mixing occurs when the exchange interaction becomes negligible compared to hyperfine coupling (L-TL region for $\VLTLeq < -0.3$~\si{\volt}), or at the \STp crossing (thin curved line of low singlet probability in Fig.~\ref{fig:spininitreadout}c). 
Theoretical calculations plotted in Fig.~\ref{fig:spininitreadout}d and experimental data are very well correlated. 
This result highlights the unique capability to control the neighboring qubit interactions \textit{in situ}. For instance, the control of the \VLTL gate voltage allows the tuning of the tunnel coupling over at least two orders of magnitude on nanosecond timescales. 
Thus, we have the ability to tailor the coupling of two adjacent dots on demand, from a spin preserving regime with a high exchange interaction to a decoupled regime, and to transfer the electron spin from the TL to the L dot.

\begin{figure}[H]
\centering
\includegraphics[width=\doublecolumn]{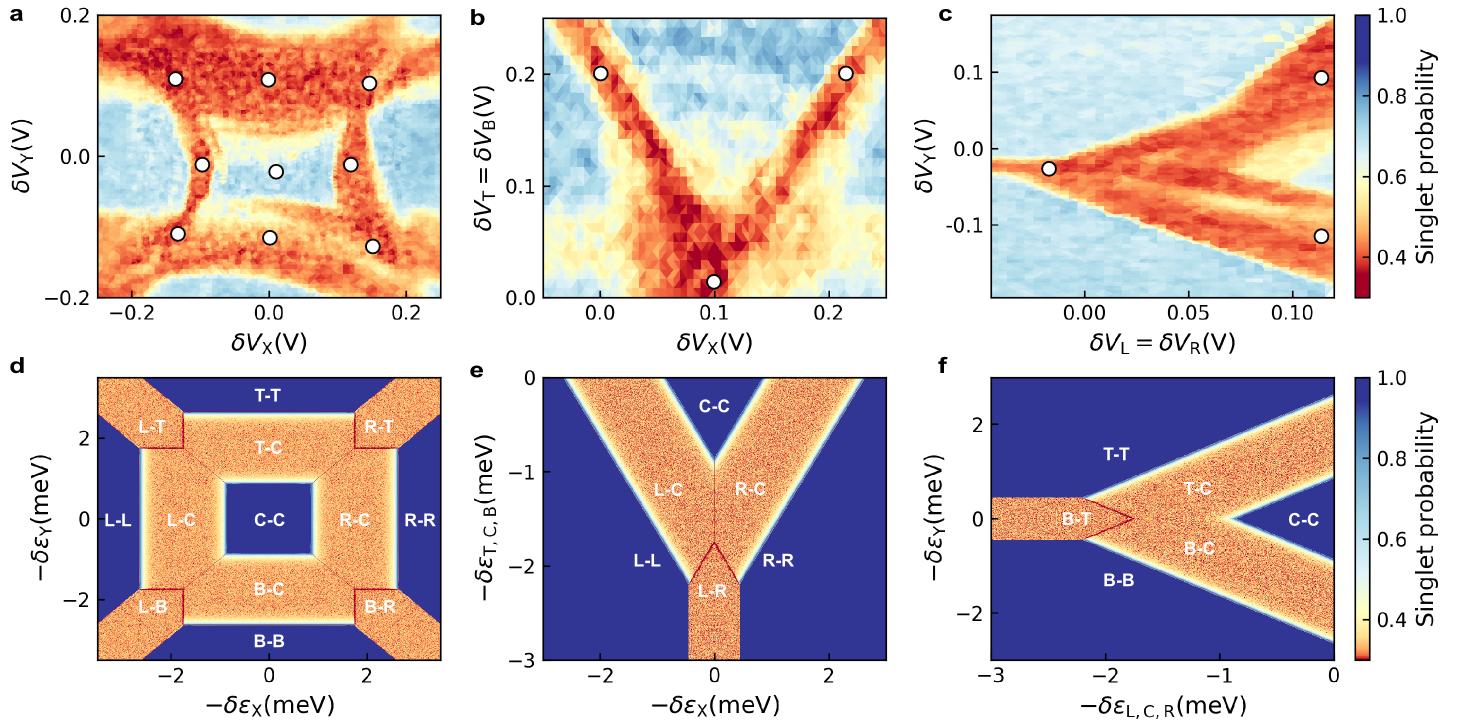}
\caption{
\textbf{Spin-mixing maps of the two-dimensional array of five quantum dots.}
\textbf{a - c,}~Two electrons in the singlet spin state initially located in C are displaced over 50~\si{\nano\second} by a voltage pulse simultaneously applied on the four \textit{orange} gates shown in Fig.~\ref{fig:sample}a. The resulting spin state is projected on the S-T$_0$ spin basis and finally read out in a single shot manner (see Methods and Extended Data Fig.~\ref{supfig:spinreadout}). The resulting singlet probability is plotted for a voltage pulse applied using ($\delta V_\text{X}$, $\delta V_\text{Y}$) in \textbf{a},  ($\delta V_\text{X}$, $\delta V_\text{T} = \delta V_\text{B}$) in \textbf{b}, and ($\delta V_\text{L} = \delta V_\text{R}$, $\delta V_\text{Y}$) in \textbf{c}. Each data point is averaged 150 times. In comparison of the spin mixing map presented in Fig.~\ref{fig:spininitreadout}, $\text{S-T}_+$ mixing line are not present and are a signature of relatively small tunnel coupling. 
Moreover, we also have indications that the interdot couplings are not symmetric, with for example a significant B-C tunnel coupling (see Extended Data Figure~\ref{supfig:DQDdecays}).
\textbf{d - f,}~Simulated spin-mixing maps realized using a homogeneous charging energy of $1$~\si{\milli\electronvolt} and tunnel coupling energy of $3$~\si{\micro\electronvolt}. The external magnetic field is set to 0 and the local magnetic fields generated by the substrate nuclear spins are assumed to follow a Gaussian distribution with standard deviation of $2.8$~\si{\milli\tesla}. 
%Simulated spin-mixing maps realized using a homogeneous charging energy $E_c = 1$~\si{\milli\electronvolt} and tunnel coupling energy $t_c = 3$~\si{\micro\electronvolt}. The external magnetic field is set to 0 and the local magnetic fields generated by the substrate nuclear spins are assumed to follow a Gaussian distribution with standard deviation of $2.8$~\si{\milli\tesla}. 
The singlet probability calculated after a 50-\si{\nano\second} time evolution is plotted as a function of ($\delta \epsilon_\text{X}$, $\delta \epsilon_\text{Y}$) in \textbf{d},  ($\delta \epsilon_\text{X}$, $\delta \epsilon_\text{T} = \delta \epsilon_\text{B}$) in \textbf{e}, and ($\delta \epsilon_\text{L} = \delta \epsilon_\text{R}$, $\delta \epsilon_\text{Y}$) in \textbf{f}.
} 
\label{fig:spinmaps}
\end{figure}

Once in the L dot, we want to demonstrate that the two electron spins can be displaced within the L-B-R-T-C cross bar geometry. 
To achieve this, we use the four gate voltages $V_\text{L, B, R, T}$ that control the four energy detunings of that subset of five QDs. 
% To achieve this, we use the four gates ($V_\text{L}$, $V_\text{B}$, $V_\text{R}$ and $V_\text{T}$) that control the four energy detunings of the subset of five QDs in a cross geometry (L, B, R, T and C). 
Similarly to Fig.~\ref{fig:spininitreadout}c, Fig.~\ref{fig:spinmaps}a represents the five QDs spin-mixing map, starting with the two electrons in the C dot and finishing in the TL dot for spin readout. 
The high singlet probability regions (blue) correspond therefore to the configurations where the electrons are in the singlet state in one of the five dots after spin preserving transfer. 
The lower singlet probability regions (red) are signature of singlet-triplet mixing after transfer with the two electrons separated in two different dots. 
Out of the $\binomialeq{5}{2}= 10$ possible separated charge configurations, only 8 are identified by comparing the experimental data with the simulation of Fig.~\ref{fig:spinmaps}d. 
The two remaining L-R or B-T configurations cannot be accessed using $\left(\dVXeq(1), \,\dVYeq(1)\right)$ (abbreviated in the following as \dVX and \dVY), but by $\left(\dVXeq(1),\,\dVYeq(-1)\right)$ (Fig.~\ref{fig:spinmaps}b, e) and $\left(\dVXeq(-1), \,\dVYeq(1)\right)$ (Fig.~\ref{fig:spinmaps}c, f), respectively. 
We therefore demonstrate that the two-electron spin can readily be initialized and read out at any position of the (TL, L, B, R, T, C) dot array.  

\subsection*{Local coherent exchange oscillations of two-electron spin states within the array}

\begin{figure}[H]
\centering
\includegraphics[width=\doublecolumn]{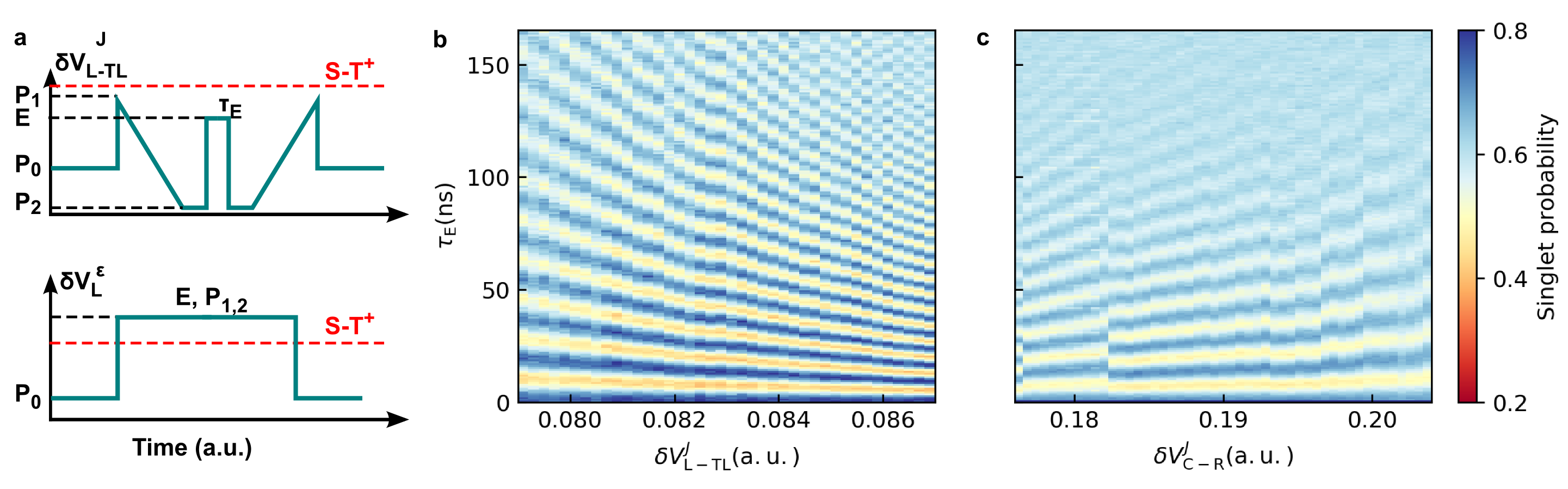}
\caption{
\textbf{Local coherent exchange oscillations in the quantum dot array} 
\textbf{a,}~Schematic $\VLTLeq(t)$ and $\VLeq(t)$ voltage sequences applied to perform the coherent exchange oscillations presented in \textbf{b}, using the voltage points defined in Fig.~\ref{fig:spininitreadout}c. 
\textbf{b,}~Coherent exchange oscillations realized with an initial spin in the TL dot and separated in TL-L charge configuration. 
The singlet probability is plotted as a function of the pulse duration  $\tau_E$ and amplitude \dJLTL. 
It is realized with a spin singlet initially in TL ($\text{P}_0$), transferred to the L-TL charge configuration ($\text{P}_1$), then adiabatically rotated to the $\ket{\uparrow\downarrow}$ spin state (slow voltage ramp from $\text{P}_1$ to $\text{P}_2$), and rapidly pulsed to E for a duration $\tau_E$.\cite{Petta2005} 
A symmetric sequence is executed to projectively readout the final spin state. We have defined for this experiment \depsilonLTL and \dJLTL so that $\depsilonLTLeq = (\dVLeq, -0.16\,\dVLTLeq)$ and $\dJLTLeq = (0.16\,\dVLeq, \dVLTLeq)$
%The voltage configurations used for this sequence are depicted in \textbf{b}.
\textbf{c,}~Similar coherent exchange oscillations realized with an initial spin in the C dot and separated in C-R charge configuration and spin transfer to the TL dot for spin readout $\left(\depsilonCReq = (-\dVLeq, \dVReq),\,\dJCReq = (\dVBeq, \dVTeq)\right)$. 
}
\label{fig:spinexchange} % fig:spininitreadout
\end{figure}

The results of charge and spin characterization presented in Figs. \ref{fig:stabdiag} and \ref{fig:spininitreadout} have been achieved with relatively moderate inter-dot couplings. 
This regime does not allow direct implementation of controlled interaction between two electron spins sitting in adjacent dots. 
However, the tunability and the fast control of the dot array nevertheless permits locally larger tunnel couplings and to reach the regime where coherent exchange of a quantum of spin between two electrons in antiparallel spin states is implemented. 
It requires, for each couple of adjacent dots, control of both tunnel coupling and detuning on fast timescales using appropriate virtual gates. 
As an example, we demonstrate it when the two electrons are located in the TL and L dots (see Fig.~\ref{fig:spinexchange}b). 
The relevant gates are then \depsilonLTL and \dJLTL (see caption of Fig.~\ref{fig:spinexchange}), and the voltage sequence and the procedure are depicted in Fig.~\ref{fig:spinexchange}{a}. 
At relatively large tunnel coupling, a change in detuning first permits the creation of a singlet state with one electron in each dot. 
The lowest energy antiparallel spin state $\ket{\uparrow\downarrow}$ with one electron in each dot is prepared by decreasing the tunnel coupling with a microsecond ramp. 
Fast branching of the exchange interaction by pulsing the tunnel coupling \cite{Bertrand2015} then allows activation of the coherent exchange interaction. 
The mirror sequence permits mapping of the $\ket{\uparrow\downarrow}$  probability onto the singlet probability measured with the spin readout procedure. 
Oscillations of the $\ket{\uparrow\downarrow}$ probability are observed as a function of the pulse duration and amplitude. 
The data points are fitted with a decaying cosine function $\cos \left( 2\pi J_\text{ex} \tau_\text{E} \right) \text{e}^{-(\tau_\text{E}/T_\text{dec})^2}$ where $J_\text{ex}$ is the exchange energy in frequency unit, and $T_\text{dec}$ the coherence time associated with the coherent exchange manipulation. 
Coherence times as long as $124 \pm 27$~\si{\nano\second} and independent of $J_\text{ex}$ are observed and consistent with the relative protection of the quantum operation pulsing the tunnel coupling.\cite{Bertrand2015, martins2016noise, reed2016reduced} 
%To demonstrate our control of the 2D array, we have implemented similar quantum manipulations for all combinations of adjacent dots in the  adjacent dots in the TL-C-T-B-L-R structure (experimental data for C-R dots shown in Fig.~\ref{fig:spinexchange}c).
Such a procedure can be implemented for any pair of adjacent dots in the array. As an example, we show in Fig.~\ref{fig:spinexchange}c the similar quantum manipulations when the electrons are positioned in the C-R dots, one of the four symmetric combinations of adjacent dots in the C-T-B-L-R structure. 
It also exhibits a coherence time independent of the exchange energy of $46\pm 8$~\si{\nano\second}. 

\subsection*{One- and two-electron coherent displacement}

To further demonstrate control of the 2D quantum dot array, we investigate the two-electron spin coherence while the electrons are continuously displaced within the C-T-B-L-R structure. 
The electrons are initially in a singlet state, and then are separated to form a coherent superposition of antisymmetric spin states. 
By shuttling one or two electrons within the dot array, the electrons will experience decoherence, resulting in the mixing with triplet states. 
The spin coherence along the displacement is probed by bringing together the electrons in the TL dot and measuring the two-electron spin states. 
Up to 10 separated charge configurations are explored, and multi-directional and complex one- and two-electron displacements are performed. 
Two timescales are relevant to understand the spin dynamics during displacement:\cite{Flentje2017-2} the first one is the so-called `displacement time', where the electrons are displaced because of the gate movement, and the second one, the `static time', where the electrons remain static in a given charge configuration. In the experiment, the displacement time is fixed and corresponds to the rise time between two voltage settings and is equal to $0.8$~\si{\nano\second}. 
We can tune the static time by changing the duration $\tau_R$ of each voltage pulses. 
In this way, we can explore the two regimes where the displacement time is either negligible or dominant compared to static time. 

\begin{figure}[H]
\centering
\includegraphics[width=\doublecolumn]{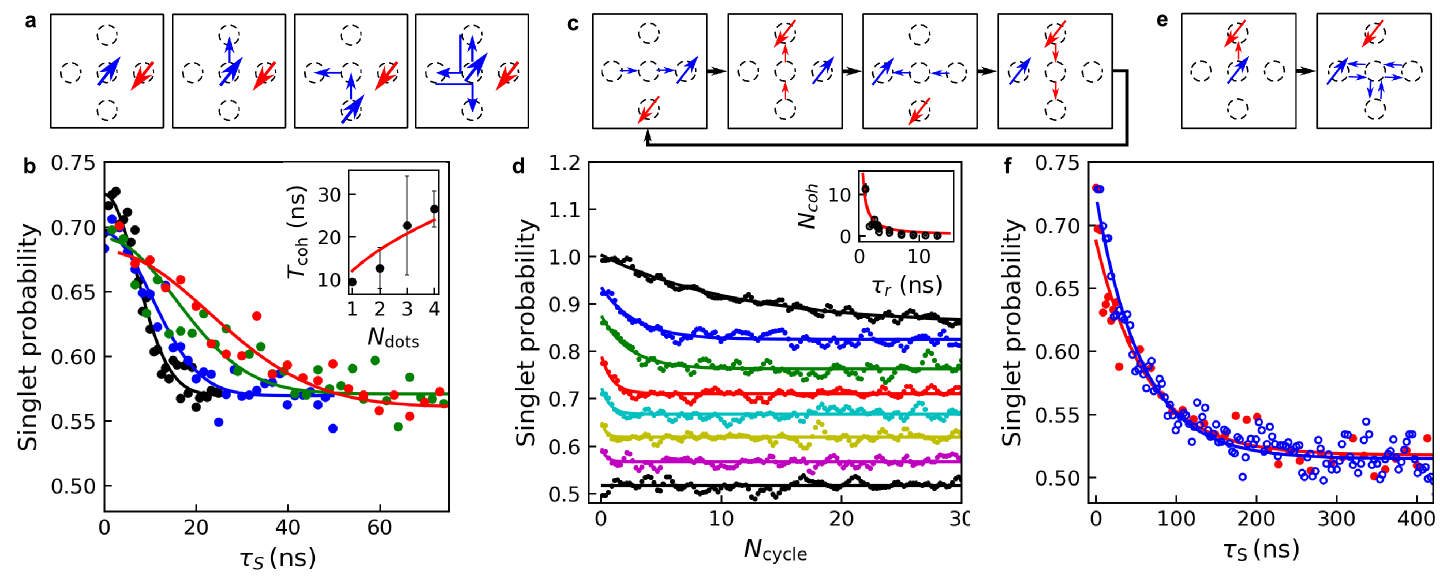}
\caption{
\textbf{Coherent electron spin displacement within a two-dimensional array.} 
\textbf{a,~c,~e,}~Schematic pictures of the different electron spin displacements performed within the two-dimensional array of five quantum dots. 
\textbf{b,}~Singlet probability plotted as a function of the total electron separation time $\tau_\text{S}$ spent in one (R-C, \textit{black}), two (R-C and R-T, \textit{blue}), three (R-B, R-C, and R-L, \textit{green}), and four (R-C, R-T, R-L, and R-B, \textit{red}) different charge configurations. The two electrons are initialized in the singlet state in the C-dot. Then, one electron is first transferred to R, and the second electron visits each other dot only once over an equal time $\tau_R$. As a consequence, the second electron spends for the different sequences a total separation time $\tau_S = 1\times \tau_R$ in C,  $\tau_S = 2\times \tau_R$ in C and T, $\tau_S = 3\times \tau_R$ in B, C, and L, or $\tau_S = 4\times \tau_R$ in C, T, L, and B, where $\tau_R$ is in integer value of the AWG clock period. The data are fitted with a Gaussian decay (\textit{solid lines}) with a characteristic time $T_2^\star$ equal to $10.0\pm0.6$, $14.6\pm1.0$, $25.5\pm2.4$, and $36.6\pm2.5$~\si{\nano\second}, respectively. 
\textbf{Inset,}~$T_2^\star$ averaged over different possible charge configurations plotted as a function $\Ndotseq$. The data are fitted with a square-root function (\textit{solid red line}). All experiments are performed under a 120-\si{\milli\tesla} magnetic field.
%\textbf{d,} Singlet probability as a function of the number of two-electron displacement cycles presented in \textbf{c} and plotted for increasing values of $\tau_\text{R}$ at 0~\si{\milli\tesla}. 
\textbf{d,}~Singlet probability as a function of the number of two-electron displacement cycles presented in \textbf{c} and plotted for  $\tau_\text{R}=1$ (black), 2 (blue), 3 (green), 5 (red), 7 (cyan), 9 (yellow), 11 (magenta), and 13~\si{\nano\second} (black) at 0~\si{\milli\tesla}. 
During one period, the electron spends $3\tau_R$ in C, and $\tau_R$ in each of the three other QDs. 
\textbf{Inset,}~Coherent number of cycles plotted against $\tau_\text{R}$. 
\textbf{f,}~Singlet probability as a function of the total electron separation time $\tau_\text{S}$ when one electron is static in R, and the second electron is periodically displaced in C, L, C, T, C, and B (blue circles), and when the two electrons $(\text{e}_1, \text{e}_2)$ are first separated in R-C, and then periodically displaced in R-B, C-B, L-B, L-C, L-T, C-T, R-T, R-C (red circles). The data points can be fitted by exponential decays, respectively characterized by the time constants of $57\pm3$ and $62\pm8$~\si{\nano\second}. 
Both experiments are realized with $\tau_\text{R} = 1.7$~\si{\nano\second}, and under a magnetic field of $120$~\si{\milli\tesla}. }
%\label{Fig5_SpinControl}
\label{fig:spincontrol}
\end{figure}

First, we investigate the situation where the displacement time is negligible compared to the static time. 
The two electrons are controllably separated in an increasing number \Ndots of separated charge configurations, each configuration being visited only one time for a time $\tau_R$ (schematics of the electron displacement shown in Fig.~\ref{fig:spincontrol}a). 
The resulting singlet probabilities are plotted (filled circles) in Fig.~\ref{fig:spincontrol}b as a function of the separated time $\tau_S = \Ndotseq \times \tau_R$. 
The coherence time is directly extracted by fitting a Gaussian decay $\text{e}^{-(\tau_S/\Tcoherenceeq)^2}$ (solid lines). 
The coherence time increases with $\sqrt{\Ndotseq}$, as the separated electrons explore a large ensemble of nuclear spins during the displacement, averaging the hyperfine interaction.\cite{Flentje2017-2, Merkulov2002, Bloembergen1948} 
We have repeated this set of experiments for different permutations of QDs in order to extract a statistical dependence of \Tcoherence on \Ndots and average out the influence of the non-regular coupling between the different dots. 
For each size of visited subsets, the average and the standard deviation of \Tcoherence is plotted in the inset of Fig.~\ref{fig:spincontrol}b as a function of \Ndots. 
They show a good agreement with the expected square root law (solid line) and confirm the successful coherent displacement within the dot array structure.

Second, we analyze the effect of the displacement time. 
In comparison with the previous electron movement, we study the situation where the electrons explore each charge configuration many times and are displaced separately, in different directions. 
To achieve this aim, we use a periodic pulse sequence engineered to maintain the two electrons separated in different QDs (see Fig.~\ref{fig:spincontrol}c): the first electron (red) is first displaced in B, and then, the second electron (blue) in R. From this point, each electron is sequentially displaced along a two-step trajectory, either vertically (red), or horizontally (blue). 
Thus, a total of eight isochronous QD configurations are periodically visited with C the only dot visited by both electrons. By changing $\tau_R$, we are able to explore the different regimes in displacement time. 
The resulting singlet probability is plotted in Fig.~\ref{fig:spincontrol}d as a function of $\tau_S = (8n+1)\tau_R$ for different values of $\tau_R$ going from $1.7$~\si{\nano\second} to $13$~\si{\nano\second}. 
The experiment has been performed at zero magnetic field. At $\tau_R$ longer than the static spin coherence time, no evolution is observed due to complete mixing between the singlet and triplet states. 
By progressively reducing $\tau_R$, we observe the emergence of a single exponential decay of the singlet probability, demonstrating preservation of the spin coherence during displacement. 
The number of coherent spin displacement cycles $N_\text{coh}$ is directly extracted from the fitting exponential decay $\text{e}^{-(N_\text{cycle}/N_\text{coh})}$ (solid lines). 
It is shown to be inversely proportional to $\tau_R$ and therefore to the speed of the electron transfer as expected for a motional narrowing process (inset of Fig.~\ref{fig:spincontrol}d and Discussion). 
For $\tau_R=1$~\si{\nano\second}, $N_\text{coh}$ and $\Tcoherenceeq$ are respectively equal to $11$ and $89$~\si{\nano\second}.
% * <matias.urdampilleta@gmail.com> 2018-06-03T10:16:00.024Z:
% 
% Define Ncycle and Ncoh
% 
% ^.

Finally, we investigate the two-electron spin coherence while only one electron is continuously displaced on multi-directional path in a two-dimensional subset of four QDs. 
We execute, periodically, a pulse sequence similar to the one described in Fig.~\ref{fig:spincontrol}a where the displaced electron is exploring the largest subset of QD (see Fig.~\ref{fig:spincontrol}e). 
We set the time spent in each charge configuration to $\tau_\text{R} = 1.7$~\si{\nano\second}. 
The experimentally measured singlet probability is plotted against $\tau_\text{S} = (6n + 1)\tau_\text{R}$ in Fig.~\ref{fig:spincontrol}f. %, where $n$ is the number of periods. 
The data point (blue circles) are fitted with an exponential decay (blue curve) $\text{e}^{-\tau_\text{S}/T_\text{flip}}$ with a characteristic time equal to $57\pm3$~\si{\nano\second}. 
We compare this two-electron spin coherence of a single  displaced electron to the case previously discussed where both electrons are displaced (red circles). % with the same conditions ($\tau_\text{R} = 1.7$). The resulting singlet probability, plotted against $\tau_\text{S} = (8n + 1)\tau_\text{R}$ (red circles), is also characterized by a single exponential decay time of $100$~\si{\nano\second}. 
It is worth noting that both electron path and number of displaced electrons are in this case different from the two-electron displacement previously discussed. 
Nevertheless, no drastic change in spin dynamics is observed between the two situations (see comparison with the red circles in Fig.~\ref{fig:spincontrol}f) as the singlet probability decays with a comparable time constant of $62\pm8$~\si{\nano\second}.

%%%%%%%%%%%%%%%%%%%%%%%%%%%%%%%%%
% Discussion
%%%%%%%%%%%%%%%%%%%%%%%%%%%%%%%%%
\section*{Discussion}

% The Discussion should be succinct and must not contain subheadings.
Three important demonstrations for the control of 2D QD arrays are presented in the manuscript: the control of the charge in a nine-dot structure with a single electron, the realization of local coherent exchange oscillations compatible with the geometry of the quantum dot array, and the coherent displacement of the electron spins within the dot array.

As far as the charge control in large QD array is concerned, a widely acknowledged problematic stands on how to probe and characterize charge stability. 
In our experiment, each relevant charge configuration could be highlighted with a specific selection of two virtual gates. 
As an example, the set of two virtual gates used in the experiment was particularly suitable for the five dots but not for the nine dots. 
Indeed, the four dots in the corner of the structure (TL, BL, TR, BR) were observed more easily with the virtual gate selection of Fig.~\ref{fig:stabdiag}d. 
The same applies with the two-electron spin maps where specific virtual gates were used to observe all the possible separated charge configurations (see Fig.~\ref{fig:spinmaps}). 
To have a complete picture, a multidimensional characterization in gate space is needed and requires the choice of a proper set of virtual gates.\cite{Hensgens2017}

The coherent exchange interaction is at the heart of the two spin-qubit gate.\cite{Brunner2011, Veldhorst2015, Zajac2018, Watson2018} 
Its implementation in the array requires a local increase of the tunnel coupling and is induced by bringing the two electrons closer. 
It is therefore based on displacing the dot potential minima over short lengths with the help of the gate voltages. 
From the perspective of large-scale coherent control, this approach is not compatible with electron spin qubits strongly coupled simultaneously with all their neighbors, and imposes a partially \textit{sequential} implementations of two-qubit gates in spin qubit arrays.

Finally, we have performed the one- and two-electron coherent displacements in a two-dimensional array. We have explored many different displacement paths through different sets of quantum dots. 
We report an increase of the coherence time which corresponds to a coherence length of five microns for both one- and two-electron displacement assuming an interdot distance of $100$~\si{\nano\meter}. 
Furthermore, we specifically study zero magnetic field regime where all the three components of the hyperfine interaction contributes equally to the decoherence of the two electron-spin singlet states. 
This hyperfine interaction is expected to be averaged along the electron displacement. 
A signature of a motional narrowing process is observed with a decoherence rate inversely proportional to the time spent in the static phase. 
Indeed, by reducing the time where the electrons are effectively static, averaging faster than the spin dynamics over many nuclear spin configurations results in an increase of the observed coherence time.\cite{Huang2013, Flentje2017} 
At the minimum static time, coherence times as long as in the non-zero magnetic regime are observed, independent of the specificity of the displacement. 
Therefore, only the single electron displacement during the gate movement from one dot to another is relevant to explain the increase in spin coherence time.
 
%%%%%%%%%%%%%%%%%%%%%%%%%%%%%%%%%
% Conclusion
%%%%%%%%%%%%%%%%%%%%%%%%%%%%%%%%%
\section*{Conclusion}
In the present manuscript, we demonstrate the control of one- and two-electron charge and spin properties in a $3\times 3$ quantum dot array. 
The tunability of the sample permits the exploration of the different charge configurations on fast timescales. 
Two-electron spin initialization, spin readout and spin transfer enable exploration of the spin dynamics in the QDs array. 
We demonstrate a procedure that permits local enhancement of the tunnel coupling between two dots of the array up to a range where coherent exchange oscillations, the basis of the two-qubit gates for electron spin qubits, are observed. 
Taking advantage of the tunability of the structure, we finally realize complex and multi-directional displacements of one and two electrons through a sub-set of five quantum dots. 
At zero magnetic field, increase of the coherence length inversely proportional to the speed of the transfer and characteristic of motional narrowing processes is reported. 
To increase even further the number of dots under control, a more scalable architecture with gates shared between many dots for control\cite{hill2015surface, li2018crossbar} and with integrated local electrometers for probing the electrons in the dot array would have to be considered.
%%%%%%%%%%%%%%%%%%%%%%%%%%%%%%%%%
% Methods
%%%%%%%%%%%%%%%%%%%%%%%%%%%%%%%%%
\section*{Methods}

% Topical subheadings are allowed. Authors must ensure that their Methods section includes adequate experimental and characterization data necessary for others in the field to reproduce their work.
\textbf{Materials and set-up.} Our device was fabricated using a Si doped AlGaAs/GaAs heterostructure grown by molecular beam epitaxy, with a two-dimensional electron gas (2DEG) 100~\si{\nano \meter} below the crystal surface which has a carrier mobility of $10^6$~\si{cm^2V^{-1}s^{-1}} and an electron density of $2.7\times10^{11}$~\si{cm^2}. 
It is anchored to the cold finger, which is in turn mechanically attached to the mixing chamber of a homemade dilution refrigerator with a base temperature of $60$~\si{mK}. 
It is placed at the center of a superconducting solenoid generating the static out-of-plane magnetic field. 
Quantum dots are defined and controlled by the application of negative voltages on Ti/Au Schottky gates deposited on the surface of the crystal. 
Homemade electronics ensure fast changes of both chemical potentials and tunnel couplings with voltage pulse rise times approaching 100~\si{\nano\second} and refreshed every 16~\si{\micro\second}. 

%below 1~\si{\micro\second}. Low temperature bias tees ensure the control of both low ($< 1$~\si{\kilo\hertz}) and high ($> 10$~\si{\mega\hertz}) frequencies of the voltages $V_\text{L, B, R, T}$. %, $V_\text{B}$, $V_\text{R}$, and $V_\text{T}$. 
Three Tektronix 5014C with a typical channel voltage rise time (20\% - 80\%) of $0.9$~\si{\nano\second} are used to rapidly change the \VL, \VB, \VR, \VT, \VLTL, \VLBL, \VBBL, \VBBR, \VRBR, \VRTR, \VTTR, and \VTTL gate voltages. 
The charge configurations can be read out by four quantum point contacts, tuned to be sensitive local electrometers, and independently biased with 300~\si{\micro\volt}. 
The resulting currents $I_\text{TL}$, $I_\text{BL}$, $I_\text{BR}$ and $I_\text{TR}$ are measured using current-to-voltage converters with a typical bandwidth of 10~\si{\kilo\hertz}. 

% Stab diag simulations
%\textbf{Stability diagram simulations.} 
\noindent\textbf{Simulations.} 
The stability diagrams are simulated using a constant interaction model. 
The total energy $U_i = Q_i^T \, C^{-1} \, Q_i$ of the one electron system is computed for each possible charge configuration and set of gate voltages ($Q_i$), assuming a realistic capacitance matrix $C$. 
The lowest energy state corresponds to the charge ground state computed for a given set of gate voltages. 
For convenience, only the charge degeneracy lines are plotted. 

% Spin map simulations
%\textbf{Spin map simulations.} 
The spin map simulations are performed with an initial singlet spin state. 
For each point in the energy detuning and tunnel coupling space, the two-electron spin Hamiltonian is computed (e.g. for double quantum dot, the Hilbert space has 12 dimensions, whereas for five quantum dots, 60 dimensions are used). 
If the energy ground state corresponds to the two electrons being in the same quantum dot, we assume that the spin singlet is conserved after a 50-\si{\nano\second} time evolution. 
On the other hand, if the two electrons are separated in different quantum dots, 
the numerical time-integration of the Schr{\"o}dinger equation is computed for 100 different hyperfine field vectors, following a centered Gaussian distribution with a standard deviation of 100~\si{\nano\electronvolt}. 
Finally, the average singlet probability is calculated.

% \bibliography{biblio}

% \noindent LaTeX formats citations and references automatically using the bibliography records in your .bib file, which you can edit via the project menu. Use the cite command for an inline citation, e.g.  \cite{Figueredo:2009dg}.

%%%%%%%%%%%%%%%%%%%%%%%%%%%%%%%%%
% Acknoledgements
%%%%%%%%%%%%%%%%%%%%%%%%%%%%%%%%%
\section*{Acknowledgements}

% Acknowledgements should be brief, and should not include thanks to anonymous referees and editors, or effusive comments. Grant or contribution numbers may be acknowledged.
We acknowledge technical support from the Poles of the Institut N\'eel, and in particular the NANOFAB team who helped with the sample realisation, as well as P. Perrier, G. Pont, H. Rodenas, E. Eyraud, D. Lepoittevin, C. Hoarau and C. Guttin. 
M.U. acknowledges the support of Marie Sklodowska Curie fellowship within the horizon 2020 european program. 
A.L. and A.D.W. acknowledge gratefully the support of 
DFG-TRR160, 
BMBF-Q.com-H 16KIS0109, 
and the DFH/UFA CDFA-05-06. 
T.M. acknowledges financial support from ERC QSPINMOTION.

%%%%%%%%%%%%%%%%%%%%%%%%%%%%%%%%%

%%%%%%%%%%%%%%%%%%%%%%%%%%%%%%%%%

%%%%%%%%%%%%%%%%%%%%%%%%%%%%%%%%%

%%%%%%%%%%%%%%%%%%%%%%%%%%%%%%%%%

%%%%%%%%%%%%%%%%%%%%%%%%%%%%%%%%%

%%%%%%%%%%%%%%%%%%%%%%%%%%%%%%%%%%
%\begin{figure}[H]
%\centering
%\includegraphics[width=\singlecolumn]{Fig6_1e_2e_Turning}
%\caption{
%\textbf{One- and two- moving electron coherent spin displacement.} 
%%
%\textbf{a,} Schematic picture of a single electron spin coherent displacement. 
%%
%\textbf{b,} Singlet probability as a function of the total electron separation time $\tau_\text{S}$ when one electron is static in R, and the second electron is periodically displaced in C, L, C, T, C, and B with $\tau_\text{R} = 1.7$~\si{\nano\second}. 
%The data are fitted with a bi-exponential function with two decay times equal to $\Tcohonee$ and $\Tfliponee$~\si{\nano\second}. 
%%
%\textbf{c,} Schematic picture of the two electron spin coherent displacement. 
%%
%\textbf{d,} Singlet probability as a function of the total electron separation time $\tau_\text{S}$ when the two electrons $(\text{e}_1, \text{e}_2)$ are first separated in R-C, and then periodically displaced in R-B, C-B, L-B, L-C, L-T, C-T, R-T, R-C with $\tau_\text{R} = 2.2$~\si{\nano\second}.
%The data are fitted with a bi-exponential function with two decay times equal to $\Tcohtwoe$ and $\Tfliptwoe$~\si{\nano\second}. 
%Both experiments are performed .  
%}
%\label{fig:1e2eturning}
%\end{figure}

%%%%%%%%%%%%%%%%%%%%%%%%%%%%%%%%%
% Extended data
%%%%%%%%%%%%%%%%%%%%%%%%%%%%%%%%%

\renewcommand{\figurename}{Extended Data Figure}
\setcounter{figure}{0}

%%%%%%%%%%%%%%%%%%%%%%%%%%%%%%%%%
\begin{figure}[H]
\centering
\includegraphics[width=\doublecolumn]{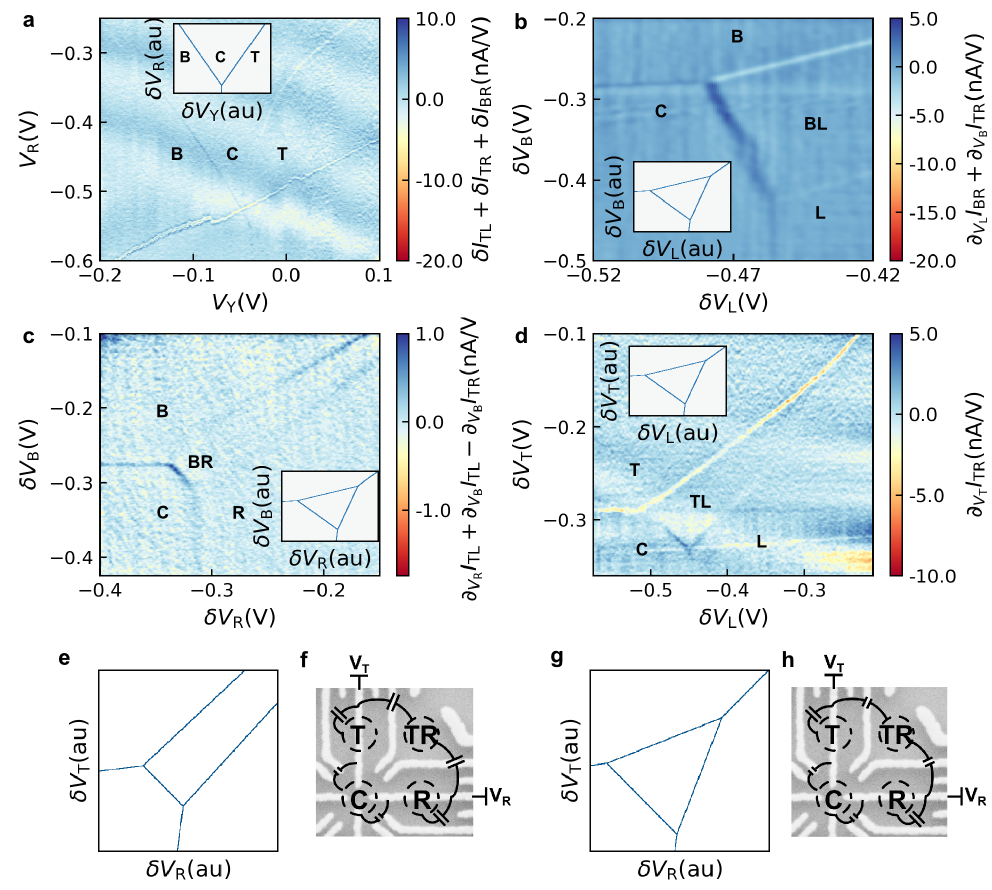}
\caption{
\textbf{One electron displacement in the $3\times 3$ array of quantum dots. } 
\textbf{a,}~Charge stability diagram of one electron in the linear triple (T, C, B) quantum dot (QD) system. 
The sum of the recorded current calculated in different directions ($\partial I_{\text{TL}}/\partial \VReq$, $\partial I_{\text{TR}}/\partial \VReq$ and $\partial I_{\text{BR}}/\partial \VReq$) is plotted as a function of the $V_\text{R}$ and $\delta V_\text{Y}(1)$. 
The labels indicate the position of the isolated electron in the QD array. 
% \textbf{b - d}, Charge stability diagrams of one electron in the four BL, L, C, B (\textbf{b}), BR, B, C, R (\textbf{c}), and TL, L, C, T (\textbf{d}).
\textbf{b,}~One electron charge stability diagram in the four BL, L, C, B QDs, recorded as a linear combination of $\partial I_{\text{BR}}/\partial \VLeq$ and $\partial I_{\text{TR}}/\partial \VBeq$ plotted as a function of $\delta V_\text{B}$ and $\delta V_\text{L}$. 
\textbf{c,}~One electron charge stability diagram in the four BR, B, C, R QDs recorded as a linear combination of $\partial I_{\text{TL}}/\partial \VReq$, $\partial I_{\text{TL}}/\partial \VBeq$ and $\partial I_{\text{TR}}/\partial \VBeq$ plotted as a function of $\delta V_\text{B}$ and $\delta V_\text{R}$. 
\textbf{d},~One electron charge stability diagram in the four  TL, L, C, T QDs recorded as $\partial I_{\text{TR}}/\partial \VTeq$ plotted as a function of $\delta V_\text{T}$ and $\delta V_\text{L}$. 
The static potential landscape is tuned in slightly different configuration for all four experiments. 
The figure insets show stability diagram simulations. 
Qualitative agreement between the experiment and the simulation topologies allow the charge configuration assignment. 
\textbf{e,}~Charge stability diagram simulation of the four T, TR, C, R QDs in the case where the capacitive coupling of the gate T to the QDs T and TR are comparable (see schematic representation \textbf{f}). In this case, the electron transfer from T to R has to be sequential via C or TR. 
\textbf{g,}~Similar charge stability diagram simulation with a significantly lower capacitive coupling between the gate T and the QD TR. Here, the electron transfer T to L may happen via a non ground state of TR (top right charge degeneracy line).
}
\label{supfig:symmstabdiag}
\end{figure}

%%%%%%%%%%%%%%%%%%%%%%%%%%%%%%%%%%
%\begin{figure}[H]
%\centering
%\includegraphics[width=\doublecolumn]{Sup_stabdiagothertopologies}
%\caption{
%Other topologies
%}
%\label{supfig:stabdiagothertopologies}
%\end{figure}

%%%%%%%%%%%%%%%%%%%%%%%%%%%%%%%%%
\begin{figure}[H]
\centering
\includegraphics[width=\singlecolumn]{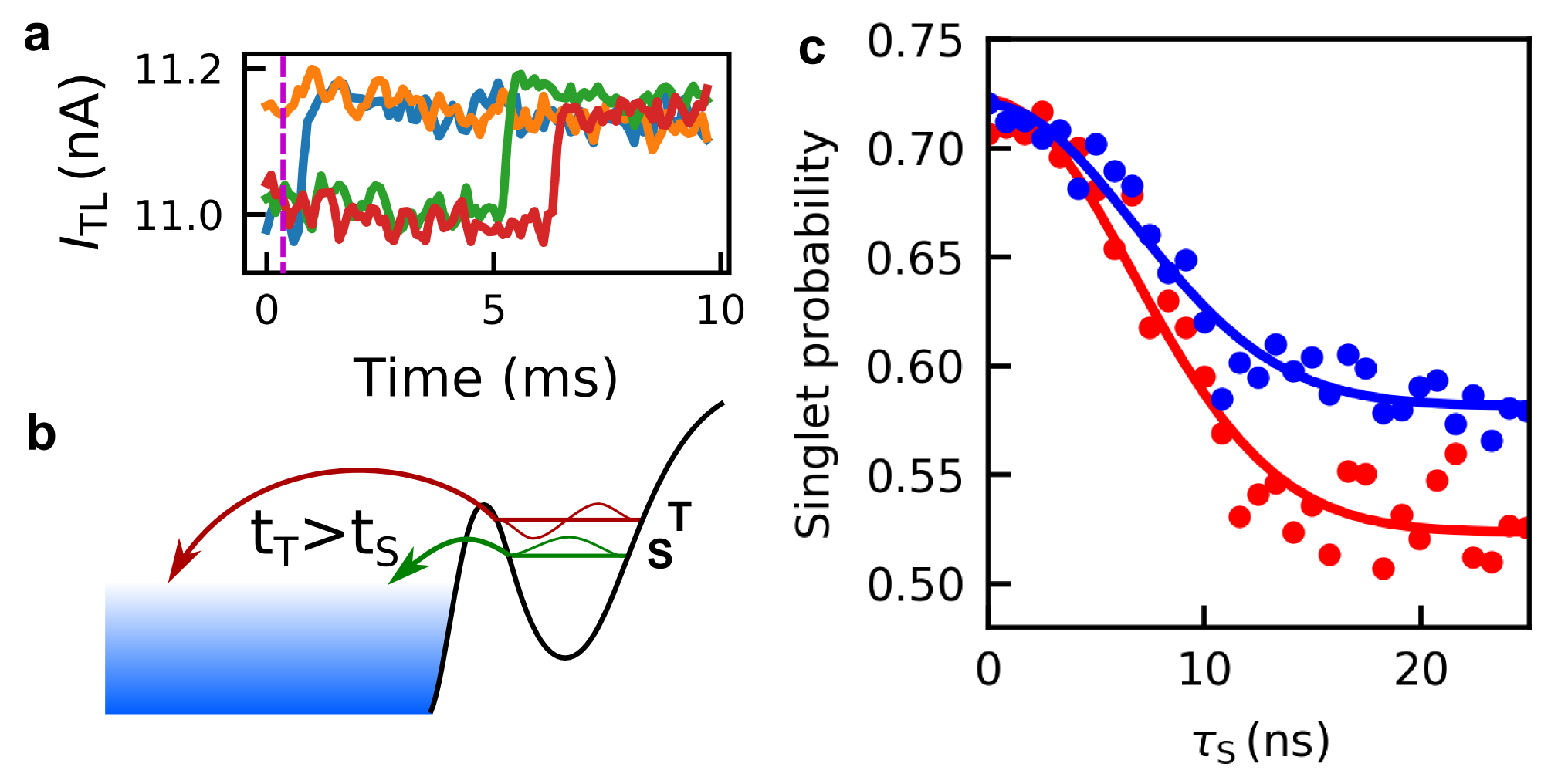}
\caption{
\textbf{Tunnel-rate dependent spin readout.}  
\textbf{a,}~Typical current traces recorded after execution of the complete pulse sequence (initialization, spin manipulation, and tuning the QD system to the readout configuration). Two current levels are assigned to either one (11.18~\si{\nano\ampere}) or two (11.0~\si{\nano\ampere}) electrons in the TL quantum dot. 
Assuming that the tunnel rate to the electron reservoir of a triplet state is much faster than for a spin singlet, counting the number of \textit{rising edges} after the time threshold (purple dashed line) directly gives access to the probability of detecting singlet states. 
\textbf{b,}~Representation of a potential cut of the TL dot. At the readout position, the charge ground state corresponds to a single electron in the dot. However, the tunnel rate of the extra electron depends on the spatial distribution of the electrons, which in turn depends on the two-electron spin state. 
\textbf{c,}~Singlet probability of a two-electron spin initially in C, and separated in C-T for a time $\tau_S$, under a 120-\si{\milli\tesla} (blue filled circles) or 0-\si{\milli\tesla} (red filled circles) out-of-plane external magnetic field. 
The initial singlet probability is 0.72, and the final probabilities are 0.58 and 0.52, respectively. 
We deduce a spin readout fidelity of 0.62.
}
\label{supfig:spinreadout}
\end{figure}

%%%%%%%%%%%%%%%%%%%%%%%%%%%%%%%%%
\begin{figure}[H]
\centering
\includegraphics[width=\singlecolumn]{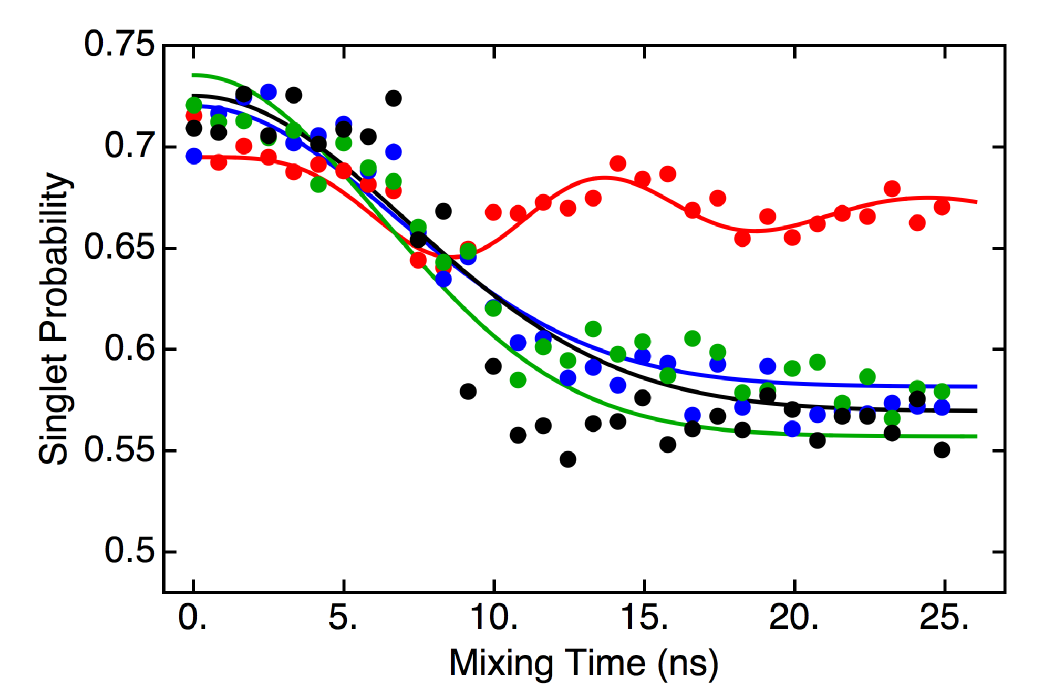}
\caption{
\textbf{Coherence decay in double quantum dots.} 
Singlet probability of a two-electron spin initially in C, and separated in either C-B (red), C-L (black), C-T (green), or C-R (blue) for a time $\tau_S$. 
We observe similar Gaussian decays (solid lines) with characteristic times of  $8.8\pm0.8$~\si{\nano\second} (C-L),  $9.5\pm0.6$~\si{\nano\second} (C-T), and  $10.0\pm0.6$~\si{\nano\second} (C-R). 
The fourth set of data points corresponding to C-B double quantum dot exhibits a non-Gaussian decaying behavior. 
Assuming a constant spin exchange interaction of 77 MHz and a standard deviation of the gradient of magnetic field in the double dot of 7.2 MHz, we calculated, using a Monte Carlo simulation, the expected time dependence of the singlet probability (purple solid curve). The similarities between the simulation and the experimental results allow us to conclude a finite spin exchange interaction is present for this configuration, and therefore a larger tunnel coupling. 
}
\label{supfig:DQDdecays}
\end{figure}

\end{document}